\documentclass[aps,prb,twocolumn,groupedaddress, showpacs,showkeys]{revtex4}

\usepackage{graphicx}
\usepackage{amsmath}
\usepackage{amsfonts}
\usepackage{bm}
\usepackage{bbm}
\usepackage{epstopdf}
\usepackage{hyperref}
\usepackage{psfrag}
\usepackage{subfigure}

\newcommand{\f}{\mathbf}

\begin{document}

\title{Multiband tight-binding theory of disordered A$_x$B$_{1-x}$C semiconductor quantum dots: Application to the optical properties of alloyed Cd$_x$Zn$_{1-x}$Se nanocrystals}

\author{Daniel Mourad}
\email{dmourad@itp.uni-bremen.de}
\affiliation{Institute for Theoretical Physics,
             University of Bremen,
             28359 Bremen, Germany}
\author{Gerd Czycholl}
\affiliation{Institute for Theoretical Physics,
             University of Bremen,
             28359 Bremen, Germany}

\date{\today}

\begin{abstract}
Zero-dimensional nanocrystals, as obtained by chemical synthesis, offer a broad range of applications, as their spectrum and thus their excitation gap can be tailored by variation of their size. Additionally, nanocrystals of the type A$_x$B$_{1-x}$C can be realized by alloying of two pure compound semiconductor materials AC and BC, which allows for a continuous tuning of their absorption and emission spectrum with the concentration $x$. We use the single-particle energies and wave functions calculated
from a multiband $sp^3$ empirical tight-binding model in combination with the configuration
interaction scheme to calculate the optical properties of 
Cd$_x$Zn$_{1-x}$Se nanocrystals with a spherical shape. In contrast to common mean-field approaches like the virtual crystal approximation (VCA), we treat the disorder on a microscopic level by taking into account a finite number of realizations for each size and concentration.
We then compare the results for the optical properties with recent experimental  data and calculate the optical bowing coefficient for further sizes.

\end{abstract}


\maketitle

\section{Introduction}

As a special realization of semiconductor quantum dots (QDs), photochemically stable semiconductor nanocrystals (NCs) can act as an alternative to organic molecules in a broad range of applications, such as light-emitting devices (LEDs),~\cite{colvin_light-emitting_1994, tessler_efficient_2002} laser applications~\cite{klimov_optical_2000} and especially for biological fluorescence labeling.~\cite{bruchez_semiconductor_1998, michalet_quantum_2005}

One common approach to directly tune the photoluminscence (PL) or absorption  spectrum of these nanocrystals in the ultraviolet-visible range is by variation of their size. This approach has extensively been used with common II-VI and III-V compound semiconductor NCs and these systems (mostly using CdSe, ZnSe, CdS and core-shell-combinations of those) are part of a large variety of experimental \cite{norris_measurement_1996,guzelian_colloidal_1996,alperson_energy_1999,xu_zinc_2007, cheng_bandgap_2009,choy_facile_2009} as well as theoretical  \cite{einevoll_confinement_1992,nair_electron_1992,ramaniah_optical_1993,wang_pseudopotential_1996,von_gruenberg_energy_1997,franceschetti_many-body_1999,lee_electron-hole_2001, lee_many-body_2002, li_first_2004,schulz_semiconductor_2007} studies.

In order to obtain emission at shorter wavelengths (500 down to 400 nm) for common CdSe NCs,  QDs with a very small diameter of less than 2 nm have to be produced,~\cite{norris_measurement_1996} which corresponds to less than 4 conventional lattice constants $a$. In this size regime, the surface passivation and size control is very difficult and one is left with a low PL efficiency.~\cite{talapin_highly_2001} When using materials with larger bulk band gaps like ZnSe or CdS, the diameter of the NCs must be enlarged and the efficiency is then lowered by the relatively weak confinement. An alternative is the usage of alloyed NCs of the type A$_x$B$_{1-x}$C with either varying or fixed size, as the emission wavelength can additionally be tuned by variation of the concentration $x$. Such alloyed NCs have for example been realized for  Cd$_x$Zn$_{1-x}$S \cite{zhong_alloyed_2003} and Cd$_x$Zn$_{1-x}$Se.~\cite{zhong_facile_2007, zhong_new_2008} Particularly, the latter Cd$_x$Zn$_{1-x}$Se NC system allows for the coverage of the whole visible spectrum with a fixed diameter of approximately 3 nm.

Similar to the behaviour of most mixed bulk semiconductors, the energetic position of the single particle gap (given by the difference $e_1-h_1$ of the lowest bound electron state and the highest hole state) as well as of the absorption/emission lines of Cd$_x$Zn$_{1-x}$Se NCs shows a pronounced bowing behaviour as a function of the concentration $x$. The deviation from linear behaviour is commonly described using a single bowing parameter $b$. The concentration dependent energetic position of the single-particle gap, respectively the spectral line  of the alloyed NCs is then given by
\begin{equation} \label{eq:bandgapbowing}
 E(x) = x \, E_{\text{AC}} + (1-x) \, E_{\text{BC}} - x \, (1-x) \, b,
\end{equation}
where $E_{\text{AC}}$ and $E_{\text{BC}}$ are the corresponding quantities of the pure AC and  BC material. In the case of binary bulk alloys, also a non-parabolic dependance of the band gap on $x$ is observed when the pure materials AC and BC have considerably different lattice constants,~\cite{richardson_dielectric_1973} which should directly carry over to the QD properties. Nevertheless, a fit of  experimental  data  by the ansatz (\ref{eq:bandgapbowing}) is still common and mostly possible. Even in the case of a perfectly parabolic behaviour, the bulk band gap $E_{\text{g}}^{\text{bulk}}(x)$, the single-particle gap $E_{\text{g}}^{\text{NC}}(x) $ and the position of the PL peak $E^{\text{PL}}(x)$ will of course not show the same bowing behaviour. 

While the single-particle states and energies themselves are not observed directly in optical measurements, their proper calculation  is of
crucial importance from a theoretical point of view. They give the foundation for a proper many-particle approach for the optical properties, like the Hartree-Fock method~\cite{williamson_multi-excitons_2001} and the configuration
interaction (CI) scheme.~\cite{williamson_multi-excitons_2001, baer_coulomb_2004}
The CI uses a finite subspace spanned by the single-particle wave functions of the QD system to construct the many-particle states.

For the calculation of the single-particle spectrum of unalloyed zero-dimensional QD systems, a broad spectrum of computational methods with different levels of sophistication can be used, like effective mass approximations,~\cite{grundmann_inas/gaas_1995, wojs_electronic_1996, shi_effects_2003} multiband $\mathbf{k}\cdot\mathbf{p}$-models,~\cite{fonoberov_excitonic_2003, pryor_eight-band_1998, stier_electronic_1999, andreev_theory_2000} empirical pseudopotential models (EPM) \cite{wang_pseudopotential_1996, franceschetti_direct_1997, franceschetti_many-body_1999,  wang_linear_1999, wang_comparison_2000, bester_cylindrically_2005} and empirical tight-binding models (ETBM).~\cite{lee_electron-hole_2001, lee_many-body_2002, santoprete_tight-binding_2003, schulz_tight-binding_2005, schulz_electronic_2006, schulz_tight-binding_2006, schulz_spin-orbit_2008}

Within the ETBM approach, one can either start from atomic orbitals localized at the sites of the crystal structure (LCAO),~\cite{slater_simplified_1954} or directly use Wannier-like orbitals localized at the sites of the Bravais lattice, i.\,e.\,on each unit cell. This defines the so-called effective bond-orbital model (EBOM).~\cite{chang_bond-orbital_1988, einevoll_effective_1990, loehr_improved_1994} As each unit cell in a  semiconductor compound of the type A$_x$B$_{1-x}$C is either occupied by the AC- or BC-Material, the spatial resolution of the EBOM is especially suitable for these mixed systems.
The basic idea in ETBM/EBOM approaches is to deduce a set of equations for the TB matrix elements in terms of a finite number of bulk parameters (taken from experiments or first-principle calculations) in order to adequately reproduce the bulk band structure. The same matrix elements then serve as input in the calculations for the nanostructures.

In order to find a suitable approximation to calculate the single-particle states for alloyed systems within a TB model, it is possible to map the microscopically disordered system on an effective system without disorder by either using a concentration-dependent interpolation of relevant parameters (virtual crystal approximation, VCA) or more sophisticated Green function methods (coherent potential approximation, CPA). The VCA is known to vastly underestimate even the bulk bowing in most cases.~\cite{boykin_approximate_2007} While the  CPA can in fact give good results for bulk crystals,~\cite{hass_electronic_1983, laufer_tight-binding_1987} it becomes much more complex and less reliable when applied to low-dimensional systems.~\cite{nikolic_coherent-potential_1992, shinozuka_electronic_2006}

An alternative for bulk as well as QD systems is to explicitely perform  calculations for an ensemble of microscopically distinct realisations, which is computationally very costful, but exact in the disorder and can reproduce bowing \cite{boykin_approximate_2007, mourad_band_2010}and line broadening effects.~\cite{oyafuso_disorder_2003}
The single-particle states can be used in CI calculations to determine the many-body states and energies and, therefore, measurable quantities like the absorption/emission spectrum. 

As it has been shown recently~\cite{mourad_band_2010}, the treatment of the EBOM with disorder on a finite supercell yields excellent results for bulk Cd$_x$Zn$_{1-x}$Se. In order to test the validity of this approach when applied to quantum dots, we  use our model to calculate the concentration-dependent emission spectrum of violet- to orange-emitting Cd$_x$Zn$_{1-x}$Se alloy nanocrystals. Such systems have experimentally been realized by means of a cation exchange reaction by Zhong \textit{et al.}\, in Ref.\,~\onlinecite{zhong_facile_2007}, and have been optically characterized by their absorption and PL spectra in the same work.

This paper is organized as follows. In Sec.\,\ref{sec:theory}, we present the theoretical approach for our calculation  of the electronic and optical properties of the Cd$_x$Zn$_{1-x}$Se alloy nanocrystals. It is based on an empirical tight-binding model with disorder on a microscopic level, and the calculation of dipole and Coulomb matrix elements from the single-particle states. Section \ref{sec:singleparticle} contains a detailed discussion of the resulting single-particle spectrum, while the following section \ref{sec:dipoleME}  discusses the influence of the disorder on the dipole and Coulomb matrix elements. In Sec.\,\ref{sec:spectrum}, we compare the resulting concentration-dependent optical spectrum to experimental results for a fixed nanocrystal size, while Sec.\,\ref{sec:calc_sizes} deals with the size-dependence of the bowing of the optical lines and the question whether the bowing parameter can be approximated by the bulk limit for larger sizes.

\section{Theory}
\label{sec:theory}
%
\subsection{Effective bond-orbital model for pure bulk semiconductors}
\label{subsec:EBOMbulk}

\begin{table}[t]
\caption{Material parameters for zincblende CdSe and
ZnSe, as used in Ref.~\onlinecite{schulz_tight-binding_2005}.}
\label{tab:param}
\begin{ruledtabular}
\begin{tabular}{llcc}
Parameter & Description &CdSe & ZnSe\\
\hline
$a$ (\AA) & lattice constant & 6.078 & 5.668\\
$E_{\mathrm{g}}$ (eV) & band gap& 1.76\footnotemark[1] & 2.82\\
$\Delta E_{\mathrm{vb}}$ (eV) & valence band offset & 0.22 & 0.00\\
$X_1^\mathrm{c}$  (eV) & X-point energy of CB & 2.94 & 4.41 \\
$X_5^\mathrm{v}$  (eV) & X-point energy of HH/LH band & -1.98 & -2.08  \\
$X_3^\mathrm{v}$  (eV) & X-point energy of SO band & -4.28 & -5.03  \\
$\Delta_{\mathrm{so}}$ (eV) & spin-orbit splitting & 0.41 & 0.43 \\
$m_\mathrm{e}$ ($m_0$) & CB effective mass & 0.120 & 0.147 \\
$\gamma_1$ & & 3.33 & 2.45  \\
$\gamma_2$  & Kohn-Luttinger parameters & 1.11 & 0.61\\
$\gamma_3$ & & 1.45 & 1.11  
\end{tabular}
\end{ruledtabular}
\footnotetext[1]{taken from Ref.\,\onlinecite{adachi_cubic_2004}.}
\end{table}

The optical properties of direct band gap semiconductors stem from an $s$-like conduction band (CB) and three $p$-like valence bands, the heavy hole (HH), light hole (LH) and split-off (SO) band. Consequently, we use one $s$- and three $p$-orbitals per spin direction, localized on the sites $\f{R}$ of the  fcc Bravais lattice underlying the zincblende crystal structure:
\begin{equation}\label{eq:sp3basis}
\left| \mathbf{R}\,\alpha \right\rangle, \quad \alpha \in \left\lbrace s\uparrow,p_x\uparrow,p_y\uparrow,p_z\uparrow,s\downarrow,p_x\downarrow,p_y\downarrow,p_z\downarrow \right\rbrace.
\end{equation}
%

The TB matrix elements of the bulk Hamiltonian $H^{\text{bulk}}$  are then given by\cite{mourad_multiband_2010}
\begin{equation}\label{eq:EBOMme}
E_{\alpha \alpha'}^{\mathbf{R} \mathbf{R'}} = \left\langle \mathbf{R} \, \alpha \right| H^{\text{bulk}} \left| \mathbf{R'} \, \alpha' \right\rangle.
\end{equation}
In order to fit the  bulk band structure to the bulk band gap $E_{\text{g}}$, the conduction band effective mass $m_e$, the spin-orbit splitting $\Delta_{\text{so}}$, the conventional lattice constant $a$, and additionally to the X-point energies of the conduction
band $X_{\text{1c}}$, the  HH and LH bands $X_{\text{5v}}$ (which are degenerate at the Brillouin zone boundary), and the SO band $X_{\text{3v}}$, we use the parametrization scheme  as presented by Loehr in Ref.\,~\onlinecite{loehr_improved_1994}. The necessary number of free parameters is achieved by the restriction of the non-vanishing hopping matrix elements up to second nearest neighbors and the usage of the two-center-approximation.~\cite{slater_simplified_1954}
The choice of material parameters for CdSe and ZnSe is presented in Tab.\,\ref{tab:param}. This set is, with one exception, identical to the one used in  Ref.\,~\onlinecite{schulz_tight-binding_2005} and has been proven to be reliable in similar TB calculations for pure CdSe and ZnSe quantum dots. It originally stems from Refs.\,\onlinecite{kim_optical_1994,  haelscher_investigation_1985}, augmented by $X$-point energies from Ref.\,\onlinecite{blachnik_numerical_1999}.

\subsection{Application to ensemble of alloyed nanocrystals}
\label{subsec:alloying}
%
\begin{figure}
\centering
\includegraphics[width=\linewidth]{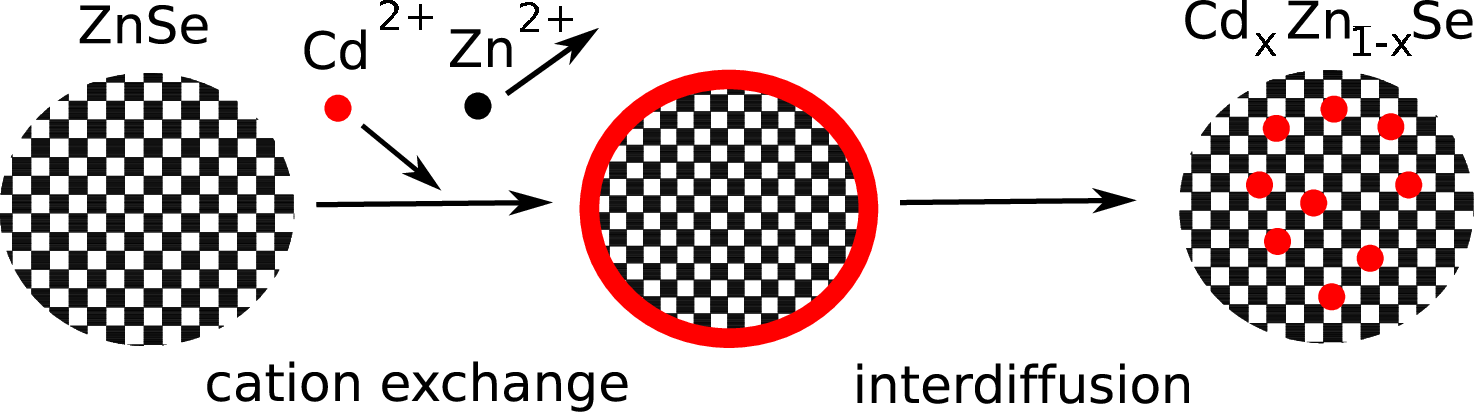}
\caption{(Color online) Realization scheme and geometry of Cd$_x$Zn$_{1-x}$Se nanocrystals by cation exchange reaction, as realized by Zhong \textit{et al.}\,(following Ref.\,\onlinecite{zhong_facile_2007}). The resulting nanocrystals have a diameter of slightly above 3 nm.}
\label{fig:expZhong}
\end{figure}

The Cd$_x$Zn$_{1-x}$Se nanocrystals from Ref.\,\onlinecite{zhong_facile_2007} are spherical in shape. Their diameter before the alloying by a cation exchange reaction is $d = (3.2 \pm 0.3)$ nm for the  CdSe NCs and $d = (3.1 \pm 0.3)$ nm for the initial ZnSe NCs, as determined by transmission electron microscopy. Their geometry and realization is schematically shown in Fig.\,\ref{fig:expZhong}.
Due to the spherical shape and their flexible surrounding, NCs without core-shell-structure show almost no strain.~\cite{alivisatos_semiconductor_1996}

In analogy to previous EBOM calculations for unalloyed zero-dimensional nanostructures,~\cite{marquardt_comparison_2008, schulz_multiband_2009, mourad_multiband_2010} we model a spherical NC with the diameter $d$, where  the localized orbital basis of  Eq.\,(\ref{eq:sp3basis}) is located on each site of the underlying fcc lattice.  As each of the sites belonging to the Cd$_x$Zn$_{1-x}$Se NC will either be occupied by the diatomic CdSe or ZnSe basis, the probability for the respective species is given by the concentrations $x$ and $1-x$ when assuming uncorrelated substitutional disorder.
Accordingly, we will use the TB matrix elements, Eq. (\ref{eq:EBOMme}), of the pure CdSe or ZnSe material for these lattice sites. For hopping matrix elements between unit cells of different material, we use the arithmetic average.

The confinement potential at the NC boundaries is assumed to be infinite and thus incorporated by setting the hopping to zero at the surface. This common  approach \cite{leung_exciton_1998, lee_electron-hole_2001, lee_many-body_2002, schulz_tight-binding_2005} corresponds to a perfect surface passivation by hydrogen or ligand
molecules.
Previous studies have shown that the effect of a finite barrier of appropriate height on the optical properties is negligible in the size regime which is considered in the present paper.~\cite{einevoll_confinement_1992}
The valence band offset (VBO) between CdSe and ZnSe is incorporated by shifting the respective site-diagonal matrix elements by the value $\Delta E_{\text{vb}} = 0.22$ eV.~\cite{schulz_tight-binding_2005} A desired finite number of single-particle electron and hole states $|\psi^{e}_i\rangle, |\psi^{h}_i\rangle$ and energies $\epsilon^{e}_i$, $\epsilon^{h}_i$  can finally be computed by numerical diagonalization of the resulting Hamilton matrix.

It should be pointed out that within the EBOM scheme, each lattice site of the A$_x$B$_{1-x}$C alloy can be unambigously assigned to the site-diagonal TB matrix elements and the corresponding material parameters  for either AC or BC. In microscopic ETBM calculations which resolve the anion and cation sites, disorder on an atomic scale prevents this, as each anion of the type C  will locally be surrounded by a different number of A or B cations.~\cite{tit_absence_2009, boykin_approximate_2007} This makes an assignment of the diagonal elements of the anions to the band structure of either AC or BC impossible  
and  will lead to an effectively more coarse-grained resolution as in the case of the EBOM.

Previous EBOM supercell calculations on bulk Cd$_{x}$Zn$_{1-x}$Se~\cite{mourad_band_2010} have shown that the bowing behaviour safely converges when the finite ensemble contains at least 50 microscopically different configurations per concentration. The same number has been found to be sufficiently reliant in the present calculations, i.\,e.\, the numerical results for the average single-particle and PL energies can be reproduced with the same degree of accuracy (10 meV) as the input parameters from Tab.\,\ref{tab:param}. In addition to the \textit{configurational} disorder, we also include the effects of \textit{concentrational} disorder,~\cite{oyafuso_disorder_2003} by not enforcing the overall concentration per configuration to equal the local probabilities for finding CdSe or ZnSe. This is crucial for the correct reproduction and prediction of experimental findings, because the experimentally determined AC or BC concentrations refer to the proportions in the ensemble and not in the single configuration.


\subsection{Coulomb and Dipole Matrix
Elements, Many-body Hamiltonian} \label{sec:ManyCoulDipGaN}

The many-body Hamiltonian in terms of creation and annihilation operators $e^{\dagger}_i$($h^{\dagger}_i$) and $e^{}_i$($h^{}_i$) for electrons
(holes) in the single-particle state $|\psi^{e}_i\rangle$
($|\psi^{h}_i\rangle$) with energy
$\epsilon^{e}_i$($\epsilon^{h}_i$) reads
\begin{equation}
H = H_0 + H_\text{C} + H_\text{D}.
\label{eq:manybodyH}
\end{equation}
It consists of the diagonal single-particle part
\begin{equation}
 H_0 = \sum_{i} \epsilon_i^e
e^{\dagger}_{i} e_{i}+\sum_{i} \epsilon_i^h h^{\dagger}_{i}
h_{i},
\end{equation}
the Coulomb interaction part
\begin{eqnarray}
H_\text{C} &=& \frac{1}{2}\sum_{ijkl} V^{ee}_{ijkl} \, e^{\dagger}_{i}
e^{\dagger}_{j} e_{k} e_{l} +\frac{1}{2}\sum_{ijkl} V^{hh}_{ijkl}\,
h^{\dagger}_{i} h^{\dagger}_{j} h_{k} h_{l} \nonumber\\
& & -\sum_{ijkl} V^{eh}_{ijkl} \, e^{\dagger}_{i} h^{\dagger}_{j} h_{k} e_{l},
\end{eqnarray}
and a part which describes the coupling to an external field  $\mathbf{E}$ in dipole
approximation:
\begin{equation}
\label{eq:lightmatter}
 H_\text{D} = \sum_{ij} \,\left(
 \langle \psi^{e}_i | e_0\mathbf{E} \cdot \mathbf{r} | \psi^{h}_j \rangle \,
e_{i}^{\phantom{\dagger}} h_{j}^{\phantom{\dagger}}\,+\,\text{h.c.}
\right),
\end{equation}
where $e_0$ is the electron charge. The present form of $H_\text{C}$ omits the contributions from electron-hole exchange terms, as they are very small compared to the other matrix elements (see Ref.\,\onlinecite{sheng_multiband_2005} for further details).
According to the discussion in Ref.\,~\onlinecite{schulz_tight-binding_2006}, the Coulomb matrix elements $V^{\lambda \lambda'}_{ijkl}$ can be
approximated by
\begin{equation}
\label{Eq:Coulappr}
 V^{\lambda \lambda'}_{ijkl}= \sum_{{\mathbf{R}}
\mathbf{R}'}\sum_{\alpha\alpha'} \left( c_{{\mathbf{R}} \alpha}^{\lambda,i} \right)^*
\left( c_{{\mathbf{R}}' \alpha'}^{\lambda',j}\right)^* c^{\lambda',k}_{{\mathbf{R}}'\alpha'}c^{\lambda,l}_{\mathbf{R} \alpha} \, V({\mathbf{R}}-{\mathbf{R}}')\,,
\end{equation}
where the asterisk denotes the complex conjugation, 
\begin{equation}
V({\mathbf{R}}-{\mathbf{R}}')=
\frac{e^2_0}{4\pi\varepsilon_0\varepsilon_\text{r}|{\mathbf{R}}-{\mathbf{R}}'|}\quad\text{for}
\quad
{\mathbf{R}}\not={\mathbf{R}}'\,\,
, \label{EqCoulappr1}
\end{equation}
and
\begin{equation}
V(0)=\frac{1}{V^2_{\text{uc}}}\int_{\text{uc}}d^3r  \, d^3r'\frac{e^2_0}{4\pi\varepsilon_0\varepsilon_\text{r}|\mathbf{r}-\mathbf{r}'|}\approx
V_0 \,.
\end{equation}
Here, $V_{\text{uc}} = a^3/4$ is the volume of the unit cell of the fcc lattice.
According to Ref.\,\onlinecite{schnell_hubbard-u_2002},
the calculation of the on-site integral $V(0)$ can be done
quasi-analytically by expansion of the Coulomb interaction in terms
of spherical harmonics.
 The  coefficients
$c^{i}_{{\mathbf{R}} \alpha}$ are numerically obtained as entries of the $i^{\text{th}}$ eigenvector of the TB matrix and are related to the corresponding one-particle wave function by
\begin{equation}
\label{eq:TBwavefunc}
\psi_i(\mathbf{r})=\sum_{{\mathbf{R}} \alpha}c^{i}_{{\mathbf{R}} \alpha}\phi_{{\mathbf{R}} \alpha}(\mathbf{r})\,\, ,
\end{equation}
where $\phi_{\mathbf{R} \alpha}(\mathbf{r})$ denotes the $\alpha$-type
effective orbital localized at the Bravais lattice site ${\mathbf{R}}$, see Eq.\,(\ref{eq:sp3basis}).
Note that for finite sized nanocrystals in solution, the proper choice of the dielectric constant $\varepsilon_\text{r}$ is  more complicated than in the case of overgrown self-assembled QD structures. We will adress this topic explicitly in the next section. 

In order to obtain the selection rules and oscillator strengths, one has to calculate the
matrix elements $\mathbf{d}^{eh}_{ij}=e_0\langle\psi^{e}_i|\mathbf{r}|\psi^{h}_j
\rangle$ of the dipole operator $e_0\mathbf{r}$ using the  EBOM wave
functions $\psi_i(\mathbf{r})$.
In the dipole Hamiltonian, Eq.~(\ref{eq:lightmatter}), we make use of the approximation~\cite{leung_electron-hole_1997, lee_many-body_2002, bryant_tight-binding_2003}
\begin{equation}
\mathbf{r} \approx \sum_{\mathbf{R} \alpha}
|\mathbf{R} \, \alpha \rangle \mathbf{R} \langle \mathbf{R}\, \alpha |,
\end{equation}
which represents an approximation for the position operator $\mathbf{r}$ in a TB approach.
Then, the TB dipole matrix elements
$d^{eh}_{ij}=\mathbf{p} \cdot \mathbf{d}^{eh}_{ij}$ explicitly read
\begin{equation}
d^{eh}_{ij}
={e_0}\sum_{\mathbf{R} \mathbf{R}'} \sum_{\alpha \alpha'} \left( c^{i,e}_{\mathbf{R} \alpha}\right)^* c^{j,h}_{\mathbf{R}' \alpha'}
\mathbf{p}\cdot{\mathbf{R}} \, \delta_{{\mathbf{R}}{\mathbf{R}}'} \delta_{\alpha\alpha'}, \label{eq:dipoleME}
\end{equation}
where $\mathbf{p} = \mathbf{E}/|\mathbf{E}|$ is the light polarization vector. Strictly speaking, the above given expressions $\mathbf{r}$ and $d^{eh}_{ij}$ give only the envelope part and neglect the spatial dependence inside the unit cells, as the latter will only give minor contributions to the overall value. We refer to Ref.\,\onlinecite{schulz_multiband_2009} for a more detailed discussion on this topic.

Even in unalloyed QD systems, band mixing effects prevent the use of selection rules based on the total angular
momentum, as it is not a good quantum number any more.~\cite{baer_influence_2007}
Additionally, the configurational disorder in our alloyed system will in general break any spatial symmetry of the NC Hamiltonian, so that each realization will show  unique  dipole and Coulomb matrix elements, which underly a certain distribution for the ensemble. In contrast to mean-field simulations, which will preserve the symmetry of the potential, this can be accounted for in our finite-ensemble approach by explicitly calculating the single-particle spectrum, dipole and Coulomb matrix elements for each concentration and configuration.


\subsection{Treatment of charge screening} \label{sec:screening}

Due to the small diameter of the NC system under consideration, finite size effects have to be taken into consideration when calulating the Coulomb matrix elements with an appropriate, site-averaged dielectric constant $\varepsilon_\text{r}$. It has been shown before that $\varepsilon_\text{r}$  will be significantly smaller than its bulk value for small QDs ($d < 5$ nm), due to a combination of finite size effects and the energy gap increase as induced by the confinement.~\cite{wang_dielectric_1994, wang_pseudopotential_1996, cartoixa_microscopic_2005} Because the Coulomb matrix elements scale directly with $1/\varepsilon_\text{r}$, a careful choice of the dielectric response is crucial for a proper reproduction and prediction of experimental spectra, especially as the Coulomb interaction  increases significantly for smaller NCs.

In the present paper, we will use a similar approach as Lee \textit{et al.} in Ref. ~\onlinecite{lee_electron-hole_2001}.  
The general idea is to use a dielectric function which is dependent on both the diameter $d$ of the quantum dot and the separation $r$ between two particles. 
The  result is a combination of the modified  Thomas-Fermi model proposed by Resta\cite{resta_thomas-fermi_1977} for the seperation dependence and a generalization of the Penn model\cite{penn_wave-number-dependent_1962} for the size dependence. It has first been used in this combination in Ref.\,\onlinecite{franceschetti_many-body_1999} for unalloyed NCs, where this approach is discussed in detail. Explicitly, this dielectric function reads
\begin{equation}
\varepsilon_\text{r}(r,d) = \begin{cases} 
\varepsilon_\infty^{\text{NC}}(d) \, q r_0 / \left[ \sinh{(q (r-r_0)) + qr} \right], & r<r_0 \\ 
\varepsilon_\infty^{\text{NC}}(d), & r \geq r_0,
\end{cases}
\label{eq:eps_rd}
\end{equation}
where
\begin{equation}
\varepsilon_\infty^{\text{NC}}(d) = 1 + (\varepsilon_\infty^{\text{bulk}} - 1) \frac{(E_{\text{g}}^{\text{bulk}} + \Delta)^2}{\left[E_{\text{g}}^{\text{NC}}(d) + \Delta\right]^2}.
\label{eq:eps_infty_NC}
\end{equation}
Here $q = (4/\pi)^{1/2}(3 \pi^2 n_0)^{1/3}$ is the Thomas-Fermi wave vector, containing the valence electron density $n_0=32/a^3$ of the zincblende structure. In our TB approach, the particles are represented by respective portions of the charge density as obtained by the $c_{\f{R} \alpha}^i$ of Eq.\,(\ref{eq:TBwavefunc}) and thus $r \approx |\f{R}-\f{R}'|$. This is on the level of a monopole approximation and consistent with the afore-mentioned treatment of the Coulomb and dipole matrix elements. $r_0$ is a critical   radius, which has to be determined by the solution of $\sinh{(q r_0)}/q r_0 = \varepsilon_\infty^{\text{NC}}(d)$ (see Ref.\,\onlinecite{franceschetti_many-body_1999}), 
$\varepsilon_\infty^{\text{bulk}}$ is the high-frequency dielectric constant of the bulk material, and $\Delta$ is basically the bulk exciton binding energy, which can be obtained from $\Delta = E_2 - E_{\text{g}}^{\text{bulk}}< 0$,  where $E_2$ is the energy of the first pronounced peak in
the bulk absorption spectrum. Following experimental data,~\cite{adachi_cubic_2004, madelung_ii-vi_1999} we will use $\Delta (\text{CdSe}) = 15$ meV and $\Delta (\text{ZnSe}) = 20$ meV  in our calculations. 

Especially for the zincblende modification of CdSe, the values for $\varepsilon_\infty^{\text{bulk}}$  differ in the literature to a certain degree, partially due to the fact that cubic CdSe is metastable and the experimental growth of high-quality samples is very difficult and often depends on the specific substrate.~\cite{mourad_band_2010} When using standard literature values, (e.\,g.\,Refs.\, \onlinecite{madelung_ii-vi_1999, adachi_cubic_2004}), this can lead to the spurious choice of $\varepsilon_\infty^{\text{bulk}}(\text{CdSe}) / \varepsilon_\infty^{\text{bulk}}(\text{ZnSe)} \approx 1$, which is not suitable for calculating the properties of the present mixed Cd$_x$Zn$_{1-x}$Se systems, as the dielectric response should become larger when the bulk band gap becomes smaller. In order to obtain a consistent parameter set, we use a version of the empirical Moss model for II-VI semiconductors,~\cite{gupta_comments_1980} where
\begin{equation}
\varepsilon_\infty^{\text{bulk}} \approx \sqrt{\frac{k}{E_{\text{g}}^{\text{bulk}}}}.
\end{equation}
In this approximation, $k$ is an empirical parameter, with $k \approx 108\, \text{eV}$ for II-VI-semiconductors.
In combination with the parameter set from Tab.\,\ref{tab:param}, this yields $\varepsilon_\infty^{\text{bulk}}(\text{CdSe}) \approx 7.83$ and $\varepsilon_\infty^{\text{bulk}}(\text{ZnSe}) \approx 6.19$. The latter is in good agreement with established experimental values for ZnSe.~\cite{madelung_ii-vi_1999}

At this point, we already anticipate that the use of the resulting single-particle gaps $E_{\text{g}}^{\text{NC}}(d)$ for different diameters $d$  of the pure or mixed NCs in Eq.\,(\ref{eq:eps_infty_NC}) always gives a screening radius $r_0$  in the range of 1.5--2.5 \AA. Thus, it does not exceed the nearest-neighbour distance of the underlying fcc lattice and according to Eq.\,(\ref{eq:eps_rd}) we will use $\varepsilon_\text{r} = \varepsilon_\infty^{\text{NC}}(d)$ in the Coulomb matrix elements $V(\mathbf{R}-\mathbf{R'})$ with $\mathbf{R}\neq\mathbf{R'}$. For the site-diagonal Coulomb interaction $V(0)$, we follow the guidelines of Ref.\,\onlinecite{lee_electron-hole_2001}. In this work, Lee \textit{et al.} performed Monte Carlo calculations with TB orbitals on similar, unalloyed Si and CdSe NCs, with the result that the effectice screening of the on-site Coulomb integrals is approximately half the long-range screening given by $\varepsilon_\infty^{\text{NC}}(d)$. In large QD systems, the resulting value of the Coulomb matrix elements is not sensitive to the exact value of $V(0)$, due to the long-ranged nature of the Coulomb interaction. We carefully checked that this statement still remains valid in the finite systems under consideration, which is also in accordance with the results of Ref.\,\onlinecite{lee_electron-hole_2001}.

\begin{figure}
\includegraphics[width=\columnwidth]{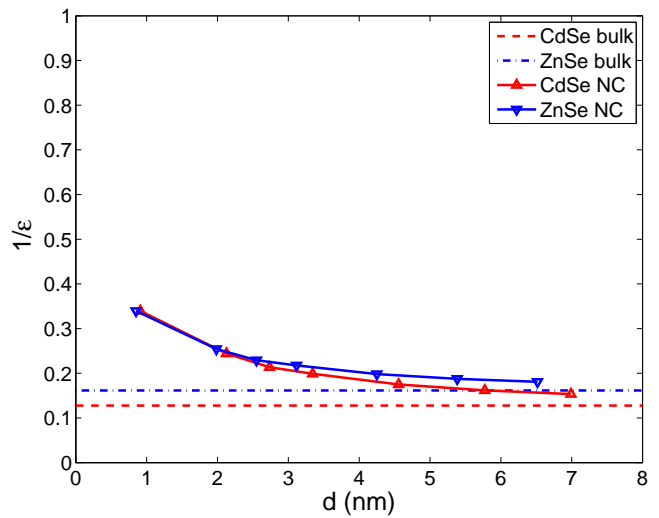}
\caption{(Color online) Resulting scaling factor $1/\varepsilon_\infty^{\text{NC}}(d)$ from the dielectric function for pure CdSe and ZnSe nanocrystals as function of the diameter $d$. For the sake of comparison, the corresponding scaling factor $1/\varepsilon_\infty^{\text{bulk}}$ for the CdSe and ZnSe bulk materials is also given.}
\label{fig:scaling_eps}
\end{figure}

For the mixed Cd$_x$Zn$_{1-x}$Se NCs, the differentiation between CdSe and ZnSe sites for the size-dependant dielectric function $\varepsilon_{\text{r}}(r,d)$ would not serve any purpose. Because any truly microscopic dielectric response function would additionally be dependent on the positions of the inducing and responding charges, the commonly used $\varepsilon_{\text{r}}$ is an effective, site-averaged quantity anyway.
The scaling factors $1/\varepsilon_\infty^{\text{NC}}(d)$ for pure CdSe and ZnSe nanocrystals as function of the  diameter $d$ are given in Fig.\,\ref{fig:scaling_eps}, together with the corresponding bulk values $1/\varepsilon_\infty^{\text{bulk}}$. As the NC scaling factor for the pure materials only differs slightly, especially for the experimentally realized size ($d \approx$ 3 nm) of Ref.\,\onlinecite{zhong_facile_2007}, we can safely use a concentrationally averaged dielectric constant,
\begin{equation}
\varepsilon_{\text{r}}(r,d,x) \approx x \, \varepsilon_{\text{r}}^{\text{CdSe}}(r,d) + (1-x) \,  \varepsilon_{\text{r}}^{\text{ZnSe}}(r,d),
\end{equation}
when calculating  the Coulomb matrix elements of the Cd$_x$Zn$_{1-x}$Se nanocrystals.

\section{Results for mixed nanocrystals}
\label{sec:results}
%
\subsection{Single-particle energies}
\label{sec:singleparticle}

%
\begin{figure*}
\includegraphics[width=\linewidth]{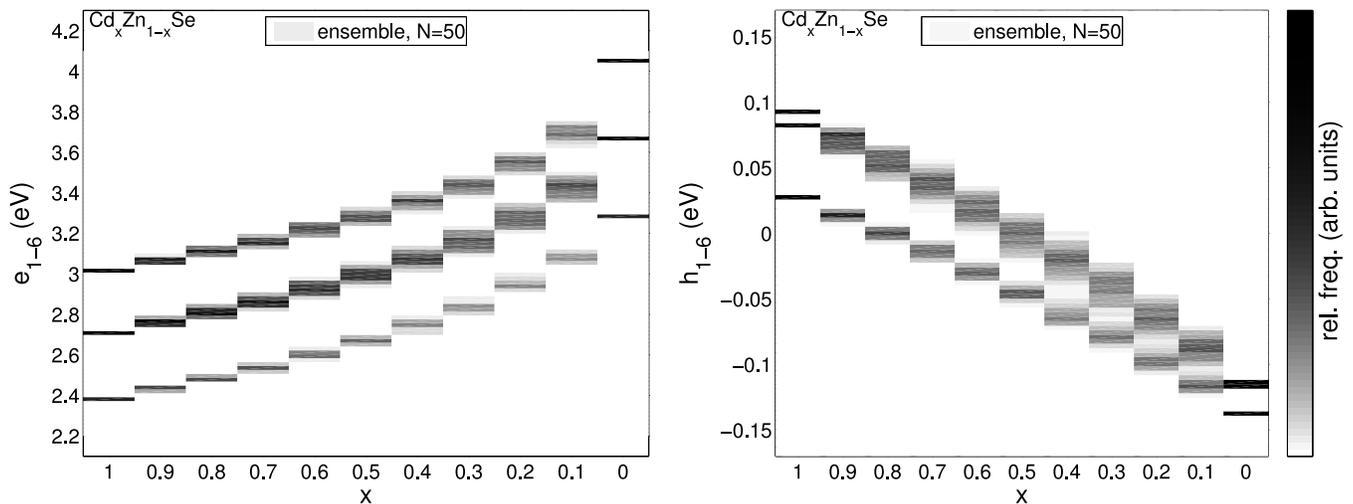}
\caption{Distribution of the single-particle energies  of the six lowest electron levels $e_{1-6}$ and six highest hole levels $h_{1-6}$ of the  Cd$_x$Zn$_{1-x}$Se nanocrystals for $N = 50$ realizations. The levels of gray give the relative frequency of the respective values. For the disordered NCs ($0 < x < 1$), the levels show a strong broadening, especially towards the Zn-richer concentrations. Each of the states $e_i$ and $h_i$ shows an additional twofold Kramers degeneracy due to time reversal symmetry. All energies are given with respect to the valence band edge of ZnSe.}
\label{fig:distr_el_and_hl}
\end{figure*}

In this section, we will discuss the single-particle spectrum of the Cd$_x$Zn$_{1-x}$Se nanocrystals as obtained from our TB calculations on a finite ensemble.
 As already discussed in detail in Section \ref{subsec:alloying}, we will assume spherical quantum dots with a diameter $d$ between $(3.1 \pm 0.3)$ nm (pure ZnSe) and $(3.2 \pm 0.3)$ nm (pure CdSe). The discretization on the fcc lattice allows for a spatial resolution of the confinement potential of half the conventional lattice constant, so that the experimental size can be modelled within the error boundaries. We account for 50 different microscopic realizations per concentration $x$, including both the effects of configurational and concentrational disorder. 


As the single-particle energies and symmetry properties of the corresponding eigenfunctions of unalloyed II-VI nanocrystals have been extensively discussed in the literature, \cite{einevoll_confinement_1992, nair_electron_1992, ramaniah_optical_1993, wang_pseudopotential_1996, franceschetti_direct_1997, franceschetti_many-body_1999, li_first_2004, schulz_tight-binding_2005, schulz_semiconductor_2007} we will here concentrate on the influence of the disorder on a finite number of eigenstates and energies.

\begin{table}
\caption{Single-particle energies of the six lowest electron levels $e_{1-6}$ and six highest hole levels $h_{1-6}$ for the unalloyed CdSe and ZnSe nanocrystals. Each of the states shows an additional twofold Kramers degeneracy. All energies are given with respect to the valence band edge of ZnSe.}
\label{tab:energies}
\begin{ruledtabular}
\begin{tabular}{lrlr}
CdSe NCs  & &ZnSe NCs  & \\
\hline
$e_1$ (eV) & $2.377$  & $e_1$ (eV)& $3.269$\\
$e_2$ (eV)& $2.703$  & $e_2$ (eV)& $3.657$\\
$e_3$ (eV)& $2.707$  & $e_3$ (eV)& $3.659$\\
$e_4$ (eV)& $2.707$  & $e_4$ (eV)& $3.659$\\
$e_5$ (eV)& $3.014$  & $e_5$ (eV)& $4.045$\\
$e_6$ (eV)& $3.014$  & $e_6$ (eV)& $4.045$ \\
\hline
$h_1$ (eV) & $0.089$  & $h_1$ (eV)& $-0.114$\\
$h_2$ (eV)& $0.089$  & $h_2$ (eV)& $-0.114$\\
$h_3$ (eV)& $0.082$  & $h_3$ (eV)& $-0.119$\\
$h_4$ (eV)& $0.082$  & $h_4$ (eV)& $-0.119$\\
$h_5$ (eV)& $0.026$  & $h_5$ (eV)& $-0.139$\\
$h_6$ (eV)& $0.026$  & $h_6$ (eV)& $-0.139$
\end{tabular}
\end{ruledtabular}
\end{table}
%

Figure \ref{fig:distr_el_and_hl} shows the distribution of the single-particle energies  of the six lowest electron levels $e_{1-6}$ and six highest hole levels $h_{1-6}$ per spin direction of the  Cd$_x$Zn$_{1-x}$Se NCs for $N = 50$ realizations, where the levels of gray give the relative frequency of the eigenvalues. Each of these levels is  twofold degenerate due to the Kramers time-reversal symmetry of the system. For the pure QD systems ($x=1$ and $x=0$), the lines show additional degeneracies that stem from the spatial symmetries of the underlying potential. The corresponding eigenenergies are additionally given in Tab.\,\ref{tab:energies} and show an excellent agreement with the results of Ref.\,\onlinecite{schulz_tight-binding_2005}, where the spectrum and its degeneracies of pure CdSe nanocrystals have been discussed by means of a fully microscopic $s_cp_a^3$ empirical tight-binding model and have furthermore been compared to experimental results by Alperson \textit{et al}.~\cite{alperson_energy_1999} Figure \ref{fig:distr_el_and_hl} also clearly shows that the energy levels of the mixed systems ($0 < x < 1$) are considerably broadened due to the microscopic disorder. The broadening is  more prominent for lower concentrations $x$ of the  small band gap material (CdSe). Especially in case of the hole levels for $x<0.5$, the disorder-induced variation of the eigenenergies exceeds the initial level splitting, so that an unambiguous assignment to the initial level order is not possible anymore. This alone renders the informative value  of the results of simple mean-field approaches like the virtual crystal approximation questionable.

\begin{figure}
\includegraphics[width=\linewidth]{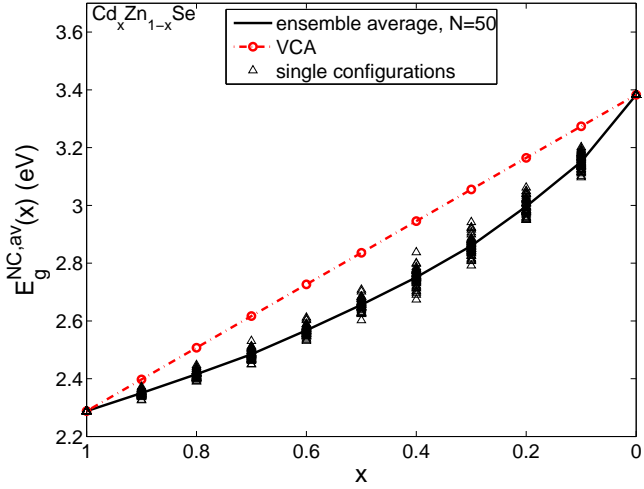}
\caption{(Color online) Average energy gap $E_{\text{g}}^{\text{NC,av}}$ for each concentration $x$ of the Cd$_x$Zn$_{1-x}$Se nanocrystals, obtained from the values depicted in Fig.\,\ref{fig:distr_el_and_hl}. The black line gives the ensemble-averaged results for $N=50$ realizations per concentration, and the red circles the VCA results. Additionally, the energy gap $ e_1^i(x) - h_1^i(x)$ of each microscopically distinct configuration is given by the black triangles.
The  VCA gives an eronneous linear behaviour over the whole concentration range. 
}
\label{fig:av_Egap}
\end{figure}

We additionally want to examine whether the virtual crystal approximation, where the TB matrix elements are obtained as a concentration-dependent linear interpolation,  can at least reproduce the behaviour of ensemble-averaged quantities and choose the single-particle energy gap $E_{\text{g}}^{\text{NC}} = e_1 - h_1$ as an example. The results are depicted in Fig.\,\ref{fig:av_Egap}. The ensemble-averaged energy gap $E_{\text{g}}^{\text{NC,av}}(x) = 1/ N \sum_{i=1}^{N} \left( e_1^i(x) - h_1^i(x) \right)$ shows a distinct bowing behaviour as a function of the concentration $x$. Clearly, the results from the VCA approach cannot reproduce any bowing behaviour and give a perfectly linear dependence of the gap on the concentration. Furthermore, even the influence of small fluctuations from the pure cases (e.\,g.\,$x=0.1$ and $x=0.9$) is vastly underestimated. For the sake of comparison, we have also added  $E_{\text{g}}^{i}(x)=  e_1^i(x) - h_1^i(x)$ for each microscopically distinct configuration, which corresponds to the energy gaps of the single nanocrystals in the ensemble.

%

In order to concentrate on the comparison with optical measurements, we will not further analyze the bowing behaviour of the single particle properties here, but proceed to calculate the excitonic spectra from the results of the TB calculations on the finite ensemble with microscopic disorder.

\subsection{Dipole and Coulomb matrix elements}
\label{sec:dipoleME}
From the single-particle wave functions, the dipole matrix elements $d^{eh}_{ij}$ and Coulomb matrix elements $V^{\lambda \lambda'}_{ijkl}$ can be calculated in the framework given in  the sections \ref{sec:ManyCoulDipGaN} and \ref{sec:screening} for each concentration and  configuration.

\begin{figure}
\includegraphics[width=\linewidth]{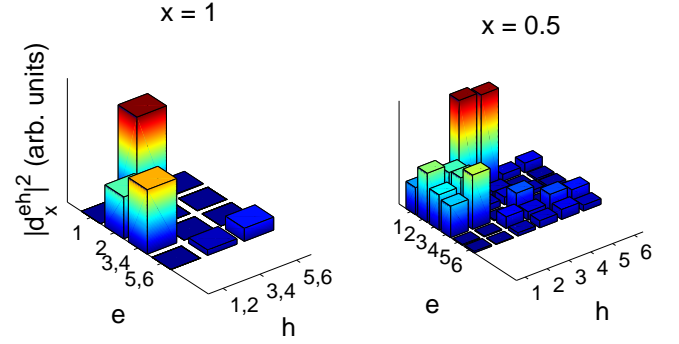}
\caption{(Color online) Ensemble-averaged modulus square of the dipole matrix element $d^{eh}_x$ for two concentrations $x$ of the Cd$_x$Zn$_{1-x}$Se nanocrystals for $N=50$ realizations (each normalized with respect to the maximum value). Degenerate states are merged into a common field. In the pure case ($x=0$), clear-cut selection rules can be obtained. Due to the loss of the spatial symmetries in the alloyed case ($x=0.5$), all degeneracies (besides the essential Kramers doublets) are lifted and the strict selection rules are relaxed.}
\label{fig:dipoleME}
\end{figure}

The ensemble-averaged modulus square of the dipole matrix element $d^{eh}_x$ for two concentrations $x$ of the Cd$_x$Zn$_{1-x}$Se nanocrystals, each normalized with respect to the maximum value, is given in Fig.\,\ref{fig:dipoleME}. Here, $d^{eh}_x$ is the projection on the (arbitrarily chosen) $\left[100\right]$ direction.
In the pure case, ($x=1$ and $x=0$), clear-cut selection rules can be obtained. The energetically lowest allowed transition for the $x=1$ and $x=0$ (not shown here) case is $e_1-h_{3,4}$, in agreement with previous results from theory and experiment.~\cite{efros_band-edge_1996, zhong_facile_2007, rajesh_exciton_2008} Due to the loss of the spatial symmetries in the alloyed case (exemplarily  shown for $x=0.5$), all degeneracies, the essential Kramers doublets left aside, are lifted and the strict selection rules are relaxed.

\begin{figure}
\includegraphics[width=\linewidth]{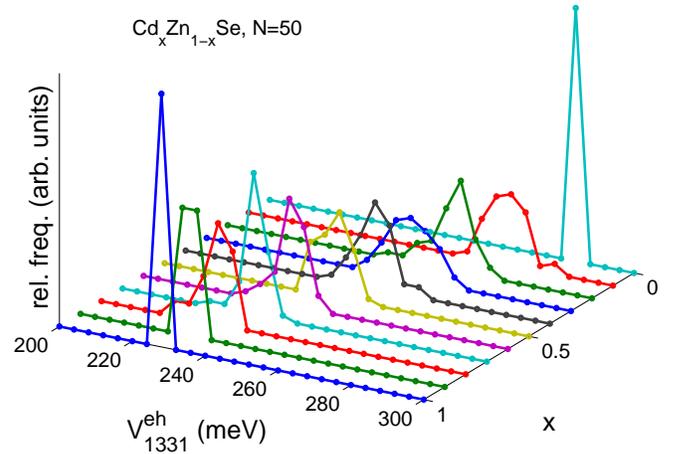}
\caption{(Color online) Distribution of the Coulomb matrix element $V^{eh}_{1331}$ for each concentration $x$ of the Cd$_x$Zn$_{1-x}$Se nanocrystals for $N=50$ realizations. In the pure cases ($x=0$ and $x=1$), this matrix element gives the main contribution to the redshift of the lowest excitonic absorption line. In the alloyed cases ($0 < x < 1$), the values show a broad distribution.}
\label{fig:CoulombME}
\end{figure}

In order to study the influence of the disorder on the Coulomb matrix element, we consequently choose the element $V^{eh}_{1331}$, which will give the main contribution to the redshift of the lowest excitonic line in the sense of a diagonal approximation (see Ref.\,\onlinecite{baer_influence_2007} for further details) in the pure case. The distribution of the values of $V^{eh}_{1331}$ for each microscopic configuration is depicted in Fig.\,\ref{fig:CoulombME} for each concentration. In the pure cases, the relative frequency is of course given by a $\delta$-distribution, as there is only one possible configuration. In the alloyed cases ($0 < x < 1$), the values show a more or less broad distribution.

\subsection{Calculation of optical spectrum  and comparison to experiment}
\label{sec:spectrum}

In order to calculate the excitonic absorption/emission spectrum for each concentration $x$ of the Cd$_x$Zn$_{1-x}$Se nanocrystals, we start from the calculated single-particle wave functions, dipole and Coulomb matrix elements for each microscopic configuration and perform CI calculations in order to obtain the many-particle states. Since we are only interested in the energetically lowest transitions, we include four electron and four hole states per spin direction. Then, the absorption/emission lines can be calculated by usage of Fermi's golden rule, 
\begin{equation}
I(\omega) = \frac{2 \pi}{\hbar} \sum\limits_{i,f} \left| \left\langle \Psi_f \right| H_{D} \left| \Psi_i \right\rangle \right|^2 \delta(E_i - E_f \pm \hbar \omega),
\label{eq:Fermi}
\end{equation} 
where $\left| \Psi_i \right\rangle$ ($\left| \Psi_f \right\rangle$) is the initial (final) many-particle state with energy $E_i$ ($E_f$), as obtained by diagonalization of the Hamiltonian (\ref{eq:manybodyH}).
For more details, we refer to  Refs.\,\onlinecite{baer_coulomb_2004} and \onlinecite{baer_influence_2007}. The resulting spectrum for the NC ensemble with concentration $x$ is then obtained by the superposition of all $N=50$ spectra.

Although there is indication that the Stokes shift in II-VI nanocrystals is either induced by the electron-hole exchange interaction,~\cite{efros_band-edge_1996, chamarro_electron-hole_2002} or exciton-acoustic phonon coupling,~\cite{takagahara_electron--phonon_1996} there also exists experimental evidence that the underlying mechanism is more complex.~\cite{liptay_anomalous_2007} As our approximation for the many-body Hamiltonian does not contain corresponding terms anyway, we will not differentiate between absorption and emission lines  in our calculations from here on. Instead, we slightly modified the TB parameters in order to reproduce the energetically lowest emission line in the pure cases ($x=1$ and $x=0$) as obtained from the experiment, which ensures the correct boundary values for the investigation of the bowing behaviour. This is the usual approach in similar	TB calculations for bulk systems,~\cite{boykin_approximate_2007, tit_absence_2009} and here necessary to eliminate further influences of e.\,g.\,the finite size resolution of both theory and experimental characterization, ambiguities in the input parameter set (especially for  CdSe, as discussed in Sec.\,\ref{sec:screening}) and the temperature dependence of the optical spectra. In retrospect, the bowing behaviour turned out to be insensitive to this procedure (not shown).

\begin{figure}
\includegraphics[width=\linewidth]{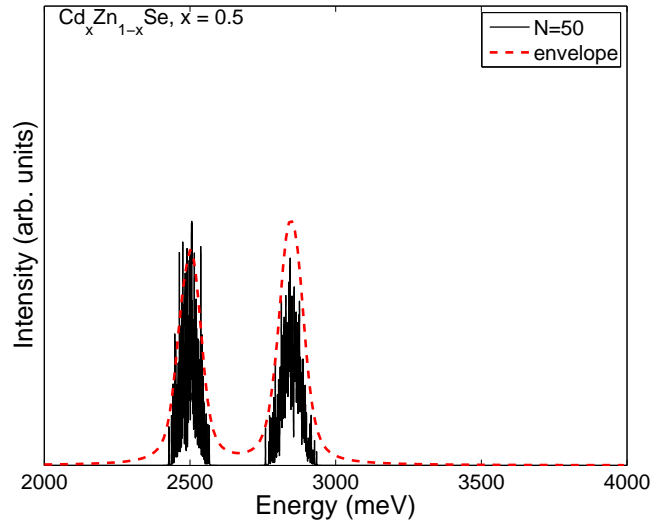}
\caption{(Color online) Optical spectrum for $x=0.5$ for the Cd$_x$Zn$_{1-x}$Se nanocrystals as obtained by the ETBM calculations on the finite ensemble. The two quasi-broadened peaks each develop from distinct, multiply degenerate  emission lines in the pure case, due to the disorder in the single-particle energies, dipole and Coulomb matrix elements.}
\label{fig:mixedspectrum}
\end{figure}

In Fig.\,\ref{fig:mixedspectrum}, we examplarily show the optical spectrum for $x=0.5$. The clearly visible two ''broadened'' peaks each develop from two distinct, multiply degenerate  emission lines in the pure case when the disorder is introduced.  Ultimately, this peak structure is  the consequence of the disorder in the single-particle energies, dipole and Coulomb matrix elements as depicted in the Figs.\, \ref{fig:distr_el_and_hl}, \ref{fig:dipoleME} and \ref{fig:CoulombME}, which introduces a different shift and line height for each configuration.  From the envelope of these broadened peaks we can read off the energetic position of the lowest  transition of the finite ensemble in dependence of $x$. The $50$ realizations per concentration turn out to be sufficient to reproducibly determine this position with an accuracy of 10 meV (which corresponds to the resolution of the input parameters, see Tab.\,\ref{tab:param}) for the NC size under consideration.

\begin{figure}
\includegraphics[width=\linewidth]{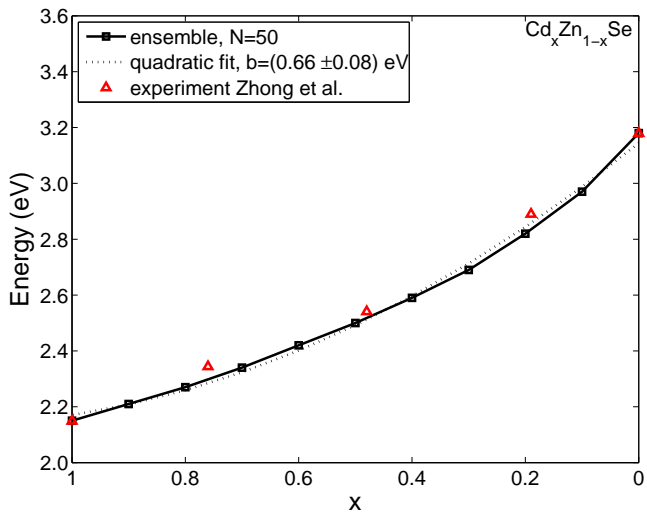}
\caption{(Color online) Energy of the lowest excitonic transition of the ensemble as a function of the concentration $x$, as obtained from the ensemble-averaged spectrum for the $d\approx 3.2$ nm Cd$_x$Zn$_{1-x}$Se NCs. The black squares give the tight-binding ensemble results , while the red triangles give
the experimental results from Zhong \textit{et al.}, as given in Ref.\, \onlinecite{zhong_facile_2007}. We calculate a bowing of $b=(0.66 \pm 0.08)$ eV, while their results give $b = (0.5\pm0.1)$ eV.}
\label{fig:comp_exp}
\end{figure}

The concentration-dependent results are summarized in Fig.\,\ref{fig:comp_exp}, where the lowest excitonic transition of the Cd$_x$Zn$_{1-x}$Se NC ensemble is given as a function of the concentration $x$. Additionally, we give the experimental results from the optical characterization by Zhong \textit{et al.}\,for the system under consideration. Obviously, there is good agreement with our theoretical curve, albeit all experimental values are slightly shifted to higher energies. We calculate a bowing of $b=0.66 \pm 0.08$ eV, while the experimental results give the smaller value $b = 0.5\pm0.1$ eV. The relatively large error shows that the common assumption of a parabolic dependence on the concentration is only of limited applicability. Nevertheless, both values agree within the error boundaries, which is especially noteworthy when comparing to bulk Cd$_x$Zn$_{1-x}$Se, where a broad range of bowing values between $b=0$  and $b=1.26$  eV is reported.~\cite{gupta_optical_1995, ammar_structural_2001, venugopal_photoluminescence_2006, tit_absence_2009} Furthermore, a comparison with similar TB-calculations and corresponding experimental results for unalloyed nanocyrstals (e.\,g.\,as in Refs.\,\onlinecite{lee_electron-hole_2001, lee_many-body_2002}) reveals the remarkably good concordance of our results with the experiment.

\subsection{Calculations for further sizes and reproduction of bulk limit}
\label{sec:calc_sizes}

%
\begin{figure}
\includegraphics[width=\linewidth]{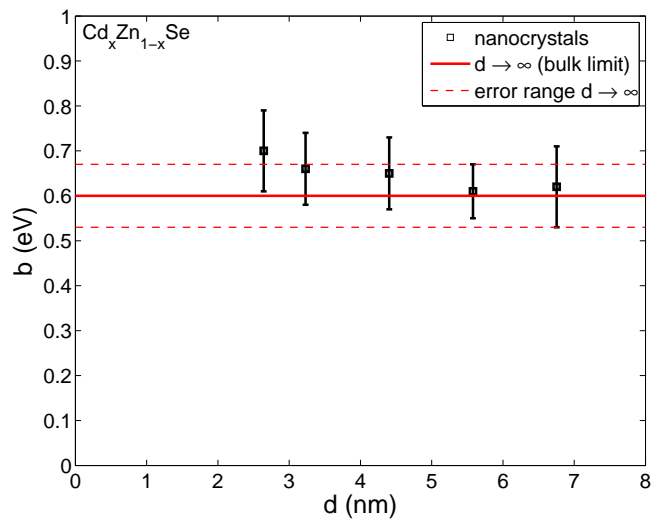}
\caption{(Color online) Size dependence of the bowing parameter $b$. Here, $d$ is the concentrationally-averaged diameter of the Cd$_x$Zn$_{1-x}$Se NCs. As before, we used $N=50$ realizations. The values approaches a bulk limit, which has been calculated with a sufficiently large supercell with periodic boundary conditions.}
\label{fig:sizedep}
\end{figure}

In this section, we finally want to discuss the dependence of the bowing parameter $b$ on the nanocrystal diameter $d$. As we found a good agreement for the Cd$_x$Zn$_{1-x}$Se nanocrystals realized by Zhong \textit{et al.} with $d \approx 3.2$ nm, we calculated the optical properties for larger sizes, where the TB formalism should be more accurate. Furthermore, we simulated the ``bulk`` limit $d\rightarrow \infty$ by using a sufficiently large supercell (4000 lattice sites) with periodic boundary conditions (see Ref.\,\onlinecite{mourad_band_2010} for more details). Again, $N=50$ realizations were used for each concentration. The results are given in Fig.\,\ref{fig:sizedep} and show that the bowing parameter reasonably approaches the bulk limit value when the diameter is increased. 

\begin{figure}
\includegraphics[width=\linewidth]{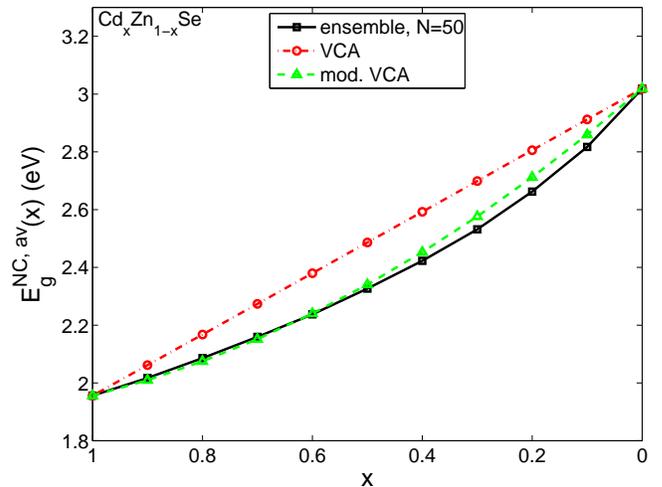}
\caption{(Color online) Average energy gap $E_{\text{g}}^{\text{NC,av}}$ for each concentration $x$ of the Cd$_x$Zn$_{1-x}$Se nanocrystals for the largest calculated size of $d \approx 6.8$ nm. The black squares give the ensemble-averaged results for $N=50$ realizations, the red circles the VCA results and the green triangles the results from the modified VCA, where the  bowing parameter of the bulk limit $d \rightarrow \infty$ has been incorporated.
The modified VCA results coincide well with the ensemble-averaged gap for the Cd-richer concentrations.}
\label{fig:av_gap_mod}
\end{figure}

Obviously, the influence of the finite size of the nanostructure on the nonlinear behaviour of the optical gap is rapidly decreasing. This raises the question, whether the bowing behaviour for larger sizes could properly be incorporated by a modified VCA approach, as suggested by di Carlo in Ref.\,\onlinecite{carlo_microscopic_2003}. Such an approach has been used in several calculations for mixed QD systems, like e.\,g.\ in Ref.\,\onlinecite{winkelnkemper_interrelation_2006} for the calculation of excitonic properties of small embedded lens-shaped In$_x$Ga$_{1-x}$N/GaN QDs  with an eight-band $\f{k}\cdot\f{p}$-model. 

Here, the diagonal TB matrix elements are weighted nonlinearly with the bulk bowing parameter, such that the correct bowing is empirically reproduced in the bulk limit. The results for the average single-particle gap $E_{\text{g}}^{\text{NC,av}}$ are depicted in Fig.\,\ref{fig:av_gap_mod}. In contrast to the linear VCA approach, the modified VCA can reproduce the average behaviour of the single-particle gap over the concentration $x$ quite well, especially for large Cd-concentrations. Nevertheless, if one is interested in the influence of small fluctuations of the small band gap material, the modified VCA still does not give accurate results, as the influence of the finite size on the energy levels and the disorder obviously  spoils the transferability of the bulk bowing to the QD energies. For smaller diameters, the deviation between the ensemble-averaged and the modified VCA results becomes more prominent (not shown).

The advantage of the modified VCA is the significantly lesser effort in the numerical calculations once the proper bulk bowing is accessible (we already discussed the broad range of the experimental bowing values, because of which we calculated the bulk bowing  ourselves within the same TB approach).  In the modified VCA case, only one calculation per concentration is necessary, so that it can surely be a feasible method if one is only interested in the concentrationally averaged properties of large systems. But although there were efforts to connect the bulk bowing  parameter $b$ to the TB parameters for the pure systems within the VCA, they fail to satisfactorily reproduce the bowing in accordance with experimental results without free parameters.~ \cite{lee_band_1990, ferhat_electronic_1996} So the main drawback of this approach remains the fact that the modified VCA  does not calculate the bowing itself but only reproduces it from an external experimental or calculated bulk bowing, in contrast to our TB calculations with microscopic disorder. Furthermore, no broadening effects can be simulated in the modified VCA. 


%

\section{Conclusion and outlook}

In this paper, we presented a  theoretical approach for the determination of the electronic and optical properties of quantum dots consisting of binary compound semiconductor alloys of the type A$_x$B$_{1-x}$C and applied it to spherical Cd$_x$Zn$_{1-x}$Se nanocrystals. 
We used a multiband empirical tight-binding approach with a basis set localized on the sites of the underlying Bravais lattice, which allows for an exact treatment of substitutional disorder on the  microscopic scale. We discussed the resulting single-particle energies of a finite ensemble of alloyed nanocrystals as a function of the concentration $x$. Furthermore, we showed by comparison to  results obtained by the virtual crystal approximation that such simple mean-field approaches cannot adequately simulate the influence of the alloying and the disorder. From the single-particle energies and wave functions, we calculated the excitonic spectrum of a finite ensemble of  50 distinct realizations per concentration by means of a configuration interaction scheme. Special attention was paid to a careful choice of the dielectric constant, in order to ensure a consistent treatment for the calculation of the Coulomb matrix elements of the mixed system.

The nonlinear concentration dependence of the spectral lines was then compared to experimental results as obtained by Zhong \textit{et al.} for NCs with a diameter of about 3 nm, yielding a very good agreement between experiment and our theory.

Finally, we presented results for larger nanocrystals and found that the bowing behaviour reasonably converges to the bulk limit. By using the bowing parameter $b$ of the bulk limit as an input parameter, one can perform a modified VCA calculation. We showed that even the modified VCA is still of only limited applicability even in the case of fairly large ($\approx$ 7 nm diameter) quantum dots.

In contrast to similar approaches for alloyed systems like Ref.\, \onlinecite{winkelnkemper_interrelation_2006}, which treat the disorder within a mean-field framework like the virtual crystal approximation, our simple TB supercell approach is able to satisfactorily reproduce the experimental findings for the alloyed system without additional free parameters.
In case that the computational resources are available, our approach can easily be applied to further alloyed QD systems, either realized by chemical synthesis or epitaxy. For systems where lattice strain plays an important role, e.\,g.\, embedded  QDs with a high lattice mismatch, a proper calculation of a strain field might have  to be additionally included for each realization, which will of course increase the calculational efforts. 

\begin{acknowledgments}
The authors would like to thank Jan-Peter Richters, Joachim Kalden, Tobias Vo{\ss} and Stefan Schulz for fruitful discussions.
We also acknowledge a grant for CPU time from the NIC at the Forschungszentrum J\"ulich and from the
Norddeutscher Verbund f\"ur Hoch- und H\"ochstleistungsrechnen
(HLRN) and would especially like to thank the consultant Thorsten Coordes for the support.  
\end{acknowledgments}


\end{document}